\documentclass[12pt]{iopart}

\usepackage{graphicx}
\usepackage{dcolumn}
\usepackage{bm}
\usepackage{rotating}

\usepackage{iopams}
\def\be{\begin{equation}}
\def\ee{\end{equation}}
\def\ba{\begin{array}}
\def\ea{\end{array}}
\def\bea{\begin{eqnarray}}
\def\eea{\end{eqnarray}}
\begin{document}

\thispagestyle{empty}
\pagenumbering{arabic}
\topmargin -1.0cm
\hoffset 0.0cm
\voffset -0.2cm
\noindent

\title[\underline{J. Phys. G: Nucl. Part. Phys. \hspace {5.9cm} T. R. Routray et al. }]
{Deformation properties  with a finite range simple effective interaction} 

\author{B. Behera$^{1\dagger}$, X. Vi\~nas$^2$, T. R. Routray$^{1*}$ L. M. Robledo$^3$, M. 
Centelles$^2$ and S. P. Pattnaik$^1$}
\address{${^1}$School of Physics, Sambalpur University, Jyotivihar-768 019, India.\\
$^2$Departament d'Estructura i Constituents de la Mat\`eria
and Institut de Ci\`encies del Cosmos (ICC),
Facultat de F\'{\i}sica, Universitat de Barcelona,
Diagonal 645, E-08028 Barcelona, Spain
\\
$^3$Departamento de F\'{\i}sica Te\'orica, Universidad Aut\'onoma de Madrid,
E-28049, Spain
\\
$^*$E-mail: trr1@rediffmail.com (corresponding author)\\
$^{\dagger}$ Retired professor}

\date{\today}

\begin{abstract}
Deformed and spherical even-even nuclei are studied using a finite range simple effective
interaction within the Hartree-Fock-Bogoliubov mean field approach. Different
parameter sets of the interaction, corresponding to different incompressibility,
are constructed by varying the exponent $\gamma$ of the density
in the traditional density-dependent term.
Ten of the twelve parameters of these interactions are determined
from properties of asymmetric nuclear matter and spin polarized pure
neutron matter. The two remaining parameters are fitted to reproduce the experimental
binding energies known in 620 even-even nuclei using several variants
of the rotational energy correction. The rms deviations for the binding
energy depend on the value of $\gamma$ and the way the rotational
energy correction is treated but they can be as low as 1.56 MeV, a 
value competitive with other renowned effective interactions of Skyrme 
and Gogny type. Charge radii are compared to the experimental values of 
313 even-even nuclei and the rms deviation is again comparable and even 
superior to the one of popular Skyrme and Gogny forces. 
Emphasis is given to the deformation properties predicted with these interactions 
by analyzing the Potential Energy Surfaces for several well 
deformed nuclei and the fission barriers of some nuclei. Comparison of the results with the experimental information, where available, as well as with the results of the Gogny D1S force shows satisfactory agreement.
\end{abstract}



\bigskip

\section{Introduction}

In a series of recent papers \cite{trr13,trr15} we have applied 
the finite range simple effective interaction (SEI) \cite{trr98,trr02},
which contains a single finite-range 
term having Gaussian form along with two $\delta$-function terms, 
to study the ground state properties 
of spherical nuclei.
At variance with typical effective interactions of Skyrme, Gogny and M3Y type,
ten of the twelve parameters of the SEI have been fitted by using 
experimental/empirical constraints in nuclear and neutron matter as discussed 
in detail in Ref. \cite{trr15}. Determination of ten out of the total twelve
parameters of SEI in nuclear matter 
allowed to reproduce the microscopic trends of the properties connected to the 
momentum dependence of the mean field and the equation of 
state (EOS) as predicted by Dirac-Brueckner-Hartree-Fock (DBHF), Brueckner-
Hartree-Fock (BHF) and variational calculations using realistic interactions 
\cite{ter87, muth00, hoffma01,samma10,dbhf3,ebhf05,pand81,Bomb91,Wu07,bal04, 
akmal01,Wiringa}. 
The two remaining parameters, one of them being the strength of the
spin-orbit interaction, are left to reproduce the binding 
energies of finite nuclei.
Pairing correlations required to describe open shell nuclei
are introduced with a density-dependent zero-range force of 
the type suggested by Bertsch and Esbenssen \cite{ber91}, 
which 
is widely used in
nuclear structure calculations \cite{gar99,sandu04,sandu05,junli08,grill11,pastore11,krewald06,baldo08}.
Within this framework
it is found that SEI can reproduce the experimental values of the binding
 energies of 161 and the charge radii of 86 even-even spherical nuclei 
with similar quality to successful Skyrme and Gogny 
effective interactions or Relativistic Mean Field models as shown in~\cite{trr15}.

In our previous works \cite{trr13,trr15}
we had used a quasi-local energy density functional (QLEDF), obtained from an Extended
Thomas-Fermi approach of the density matrix \cite{vinas00,vinas03}, plus a BCS
treatment of pairing correlations to study the ground-state properties of 
spherical nuclei. Our aim in this work is to extend the description of finite nuclei with the 
SEI by performing full Hartree-Fock-Bogoliubov (HFB) calculations 
including
deformation degrees of freedom. 
The paper is organized as follows. In section 2 the formalism
concerning nuclear matter and finite nuclei using the SEI is briefly
summarized. In the third section the
predictions for the binding energies of 620 deformed and spherical even-even nuclei
across the nuclear chart with known experimental
masses \cite{AW.12}, are made in the HFB
framework for the different sets of the SEI interaction.
The charge radii of 313 even-even nuclei
predicted by these sets of SEI are also compared with
the experimental data \cite{angeli04}.
Comparison of the results of binding energies and charge radii for
even-even spherical nuclei computed with the QLEDF
used in
Refs.~\cite{trr13,trr15} with the corresponding
HFB results of the present work
is also given in this section.
The fourth section is devoted to study the SEI predictions in some of the
problems crucially dependent on deformation properties.
In particular, we analyze the Potential Energy Surfaces (PES) of some
typical nuclei and the fission barriers of
$^{240}$Pu and $^{256}$Fm. The
predictions are compared with the experimental values where available
and with the results obtained with the Gogny D1S force, which can
be considered as a benchmark in the theoretical 
study of deformed nuclei and fission barriers \cite {blaizot95}. Finally our conclusions are given in the last section.
  
\section{Formalism}

The finite range simple effective interaction used to describe nuclear matter and finite nuclei
is given by
\begin{eqnarray}
v_{eff}({\bf r},{\bf R})&=&t_0 (1+x_0P_{\sigma})\delta({\bf r}) \nonumber \\
&&+\frac{t_3}{6}(1+x_3 P_{\sigma})\left(\frac{\rho({\bf R})}
{1+b\rho({\bf R})}\right)^{\gamma} \delta({\bf r}) \nonumber \\
&& + \left(W+BP_{\sigma}-HP_{\tau}-MP_{\sigma}P_{\tau}\right)e^{-r^2/\alpha^2},
\label{eq1}
\end{eqnarray}
where, ${\bf r}$ and ${\bf R}$ are the relative and center of mass coordinates,
respectively. The SEI in equation (\ref{eq1}) has in total 12 parameters,
 namely, 
$b$, $t_0$, $x_0$, $t_3$, $x_3$, $\gamma$, $\alpha$,
$W$, $B$, $H$ and $M$ plus the spin-orbit strength parameter $W_0$, which enters
in the description of finite nuclei. The SEI interaction is similar in 
form to the Skyrme force where the gradient terms are replaced by the 
single finite-range contribution. A similar analogy can be drawn with the 
Gogny interaction where one of the two finite range terms is replaced by the 
zero-range $t_0$-term. One more difference in this context is that the density-dependent
term of SEI contains the factor (1+b$\rho$)$^\gamma$ in the denominator, 
where the parameter $b$ is fixed to prevent the supra-luminous behavior in nuclear matter
at high densities \cite{trr97}.
The formulation of nuclear matter and neutron matter using SEI has been 
discussed at length in Refs.~\cite{trr13,trr15}, but for the sake of convenience of the 
reader we report in the Appendices A and B the expressions of the energy densities 
of asymmetric nuclear matter, spin polarized neutron matter and finite nuclei 
obtained with SEI given in equation (\ref{eq1}). 
We shall now outline, in brief, the determination of the parameters 
involved in the study of nuclear and neutron matter. The study of asymmetric 
nuclear matter involves altogether nine parameters, 
namely, $\gamma$, $b$, $\varepsilon_{0}^{l}$, $\varepsilon_{0}^{ul}$,
$\varepsilon_{\gamma}^{l}$,$\varepsilon_{\gamma}^{ul}$, $\varepsilon_{ex}^{l}$,
$\varepsilon_{ex}^{ul}$ and $\alpha$, whose connection to the interaction 
parameters is given in the Appendix A. 
However, symmetric nuclear matter requires only the following combinations
of the strength parameters in the like and unlike channels
\begin{eqnarray}
\left(\frac{\varepsilon_{0}^{l}+\varepsilon_{0}^{ul}}{2}\right)=\varepsilon_0,
\left(\frac{\varepsilon_{\gamma}^{l}+\varepsilon_{\gamma}^{ul}}{2}\right)=\varepsilon_{\gamma},
\left(\frac{\varepsilon_{ex}^{l}+\varepsilon_{ex}^{ul}}{2}\right)=\varepsilon_{ex},
\label{eq9}
\end{eqnarray}
together with $\gamma$, $b$ and $\alpha$, 
i.e., altogether six parameters.

For a given value of the exponent
$\gamma$, and assuming 
the standard values for nucleon mass, saturation density and 
binding energy per particle at saturation, 
the remaining five parameters $\varepsilon_{0}$, $\varepsilon_{\gamma}$,
$\varepsilon_{ex}$, $b$ and $\alpha$ 
of symmetric nuclear matter are determined in the following way. 
The range $\alpha$ and the exchange strength $\varepsilon_{ex}$ are determined 
simultaneously by adopting an optimization procedure using the condition 
that the nuclear mean field in symmetric nuclear matter at saturation density 
vanishes for a kinetic energy of the nucleon of 300 MeV, a result extracted from optical model analysis
of nucleon-nucleus data \cite{gale87,gale90,cser92}. 
The parameter $b$ is determined as mentioned before.
The two remaining parameters,
namely $\varepsilon_{\gamma}$ and $\varepsilon_{0}$, are obtained from the saturation
conditions. The stiffness parameter $\gamma$ is kept as a free parameter 
and its allowed values are chosen in such a way that the corresponding pressure-density relation in symmetric matter lies within the region compatible 
with the analysis of flow data in heavy-ion collision experiments 
\cite{Danielz02}. It is found that the maximum value that fulfills this condition is $\gamma$=1, which corresponds to a nuclear matter incompressibility 
$K(\rho_0)$=283 MeV. Therefore, we can study the 
nuclear matter properties by assuming different values of $\gamma$ up to 
the limit 
$\gamma$=1. Now,
to describe asymmetric nuclear matter we need to know how the strength 
parameters
$\varepsilon_{ex}$, $\varepsilon_{\gamma}$ and $\varepsilon_{0}$ 
of equation (\ref{eq9})
split into
the like and unlike components. The splitting of $\varepsilon_{ex}$ into
$\varepsilon_{ex}^{l}$ and $\varepsilon_{ex}^{ul}$ is decided from the 
condition that the entropy density in pure neutron matter should not exceed 
that of the symmetric nuclear matter. This prescribes the critical value for 
the splitting of the exchange strength parameter to be $\varepsilon_{ex}^{l}=2\varepsilon_{ex}/3$ \cite{trr11}. 
The splitting of the remaining two strength parameters 
$\varepsilon_{\gamma}$ and $\varepsilon_{0}$,
is obtained from the values of the symmetry energy $E_s(\rho_0)$ and 
its derivative $E_s^{'}(\rho_0)$ = $\rho_0 \frac{dE_s(\rho_0)}{d\rho_0}$ 
at saturation density $\rho_0$. By assuming a value for 
$E_s(\rho_0)$ within its accepted range \cite{dutra12}, 
we determine $E_s^{'}(\rho_0)$ from the condition that the difference 
between the energy densities of the nucleonic part in charge neutral 
beta-stable $n+p+e+\mu$ matter and in symmetric matter at the same density 
be maximal \cite{trr07}. The value of $E_s^{'}(\rho_0)$ thus obtained 
predicts a density dependence of the symmetry energy which is neither very stiff
nor soft. With the parameters 
determined in this way, the SEI was able to reproduce 
the trends of
the EOS and the momentum dependence of the mean field properties 
with similar quality as predicted by microscopic calculations
\cite{trr07,trr11}. We have still two free parameters that we have taken
to be $t_0$ and $x_0$.
 In our first work on finite nuclei in Ref.~\cite{trr13},
the $t_0$ and $x_0$ parameters, along with the spin-orbit strength $W_0$,
were determined by a simultaneous fit to the experimental binding energies
of $^{40}$Ca and $^{208}$Pb and to the splitting of the $1p$ single-particle levels in $^{16}$O.
With the parameterizations corresponding to $\gamma=1/3 \, (1/2)$, the SEI predicted  {\it rms}
deviations in 161 binding energies and 86 charge radii of even-even spherical 
nuclei of 1.39\,(1.54) MeV and 0.017\,(0.015) fm, respectively.

In our subsequent work 
in  Ref.~\cite{trr15}, we analyzed in detail the predictions of the SEI in the spin channel.
It was found that the $x_0$ and $W_0$ parameters are, actually, correlated. Different combinations of $x_0$ and $W_0$ leave the {\it rms} deviations of binding energies and charge radii practically unchanged,
but predict a very different behavior in spin polarized matter which is 
sensitive to the value of $x_0$. To determine the $x_0$ parameter uniquely, we considered the
particular case of spin polarized neutron matter. 
Its description requires to know how the strength parameters $\varepsilon_{0}^{l}$, $\varepsilon_{\gamma}^{l}$ and
$\varepsilon_{ex}^{l}$ of spin saturated neutron matter split into the two channels of like (parallel, ``l,l") and 
unlike (anti-parallel, ``l,ul") spin orientations under the constraints
$\varepsilon_{0}^{l}$=$(\varepsilon_{0}^{l,l}+\varepsilon_{0}^{l,ul})$/2,
$\varepsilon_{\gamma}^{l}$=$(\varepsilon_{\gamma}^{l,l}+\varepsilon_{\gamma}^{l,ul})$/2 
and $\varepsilon_{ex}^{l}$=$(\varepsilon_{ex}^{l,l}+\varepsilon_{ex}^{l,ul})/2$.
The completely polarized neutron matter computed with the SEI imposes that  
${\varepsilon_{\gamma}^{l,l}}$=0 and ${\varepsilon_{0}^{l,l}}=-{\varepsilon_{ex}^{l,l}}$. 
Further, from a fit to the DBHF results on the effective mass splitting between spin-up and spin-down neutrons 
in spin polarized neutron matter, it is found that the SEI predictions agree well with the DBHF ones 
for $\varepsilon_{ex}^{l,l}=\varepsilon_{ex}^{l}/3$. 
This consideration allows to determine {$x_0$} in a unique way as \cite{trr15},
\begin{equation}
x_0 = 1 - \frac{2\varepsilon_{0}^{l} - \varepsilon_{ex}^{l}}{\rho_0 t_0},
\label{eq9a}
\end{equation}
assuming that $t_0$ is known. 
 The two remaining free parameters, $t_0$ and $W_0$, were
fitted in \cite{trr15} to reproduce the binding energies of $^{40}$Ca and $^{208}$Pb, respectively. 
The values of the nine parameters of the SEI, which fully describe the asymmetric nuclear
matter for $\gamma=1/3$ and $\gamma=1/2$, are presented in Table 1. The nuclear matter 
saturation properties, such as, saturation density $\rho_0$, energy per particle $e(\rho_0)$, 
incompressibility $K(\rho_0)$, effective mass $m^{*}/m$, symmetry energy $E_{s}(\rho_0)$ and slope parameter  of the symmetry energy $L(\rho_0)$ for these sets are listed in Table 2.
The values of  $t_0$ and $W_0$, obtained from the fit to the binding energies of $^{40}$Ca and $^{208}$Pb,
along with the values of $x_0$ provided by equation (\ref{eq9a})
for the SEI with  $\gamma=1/3$ and $\gamma=1/2$ are 
given in the two first lines of Table 3 (see below).
\begin{table*} \caption{Values of the parameters of asymmetric nuclear matter for the two EOSs corresponding to 
$\gamma$=1/3 and 1/2 used in this work.}

\begin{tabular}{ccccccccc}\hline \hline
$\gamma$ & $b$&$\alpha$&$\varepsilon_{ex}$&$\varepsilon_{ex}^{l}$&$\varepsilon_{0}$&${\varepsilon_{0}}^{l}$&
$\varepsilon_{\gamma}$&$\varepsilon_{\gamma}^{l}$\\
&   fm$^{3}$ &fm&MeV&MeV&MeV&MeV&MeV&MeV \\ \hline
1/3&0.4184&0.7582&-95.6480&-63.7653&-112.7493&-67.0819&110.7436&78.7768 \\
1/2&0.5914&0.7597&-94.4614&-62.9743&-78.7832&-45.8788&77.5068&57.7687 \\
\hline \hline
\end{tabular}
\end{table*}

\begin{table*} \caption{Values of nuclear matter properties at saturation
for the two EOSs corresponding to $\gamma$=1/3 and 1/2. }
\begin{tabular}{ccccccc}\hline \hline
\centering
$\gamma$ &
$\rho_0(\mathrm{fm}^{-3})$&$e(\rho_0)(\mathrm{MeV})$&$K(\rho_0)(\mathrm{MeV})$&$\frac{m^*}{m}(\rho_0,k_{f_0})$
&$E_{s}(\rho_0)(\mathrm{MeV})$&$L(\rho_0)(\mathrm{MeV})$\\
1/3&0.1597&-16.0&226&0.710&35.0&74.94\\
1/2&0.1571&-16.0&245&0.711&35.0&76.26\\
\hline \hline
\end{tabular}
\end{table*}

In the present work, we study deformed nuclei with the SEI and to this end 
we perform calculations in the mean field approach using the HFB 
method. These calculations are restricted to axially symmetric 
geometry. The pairing interaction is the same as in 
Refs.\cite{trr13,trr15}. The solution of the HFB equations has been 
recast as a minimization procedure of the energy density functional 
(see Appendix B), where the HFB wave function of the Bogoliubov 
transformation is chosen to minimize the energy. The quasiparticle 
operators entering in this transformation are expanded in a harmonic 
oscillator basis with a number of shells depending of the nucleus 
considered.  An approximated second-order gradient method \cite{robledo11}
is used to solve the non-linear HFB equations. Once the minimal
energy solution is found, an extrapolation of the energy to an infinite
number of shells \cite{hilaire07} is performed (see also \cite{BCPM13} 
in this respect). 

\section{Results}

\subsection{Binding energies and charge radii of finite nuclei}

The parameters $t_0$  and $W_0$ of the SEI are fixed as to reproduce 
 the experimental binding energies of known  even-even finite nuclei. The
mathematical procedure is to minimize the binding energy {\it rms} 
deviation $\sigma(E)$ defined as
\begin{equation}
\sigma^2(E) = \frac{1}{N} \sum_{i=1}^N (B_{th}(i) - B_{exp}(i))^2,
\label{eq9b}
\end{equation}
where the sum extends to the 620 even-even nuclei with known 
experimental masses \cite{AW.12}. For the theoretical binding energies 
we will consider three different possibilities. The first one is
the value given by the HFB energy of the ground state minimum.
The second possibility is the HFB binding energy supplemented by the rotational energy correction $E_\mathrm{rot}$
based on the HFB ground state. It
corresponds to a projection after variation (PAV) procedure where the
projected energy is computed using the rotational formula approximation
and the HFB ground state wave function-- see \cite{egido04,RRG00} for details.
Note that the rotational energy
correction plays an important role in deformed nuclei and its inclusion is
relevant to describe masses along the whole periodic table.
In strongly deformed mid-shell heavy nuclei the rotational energy 
correction can reach values as large as 6 or 7 MeV. This correction, however, is 
almost negligible in magic or semi-magic nuclei, which are basically 
spherical. Due to the fact that the spherical-deformed transition is sharp,
the rotational correction goes from zero to some MeV at the transition point leading
to sharp variations in the binding energy plot. In order to find a remedy to
this drawback 
we consider also a restricted variation after projection (RVAP), where 
the approximate projected energy is minimized in the space of HFB wave 
functions constrained to given values of the quadrupole moment. In this 
way, a relatively smooth behavior in the binding energy plot (see below) is 
obtained.  

\begin{sidewaysfigure}
\vspace*{15cm}
\centering
\includegraphics[width=0.45\columnwidth,angle=-90]{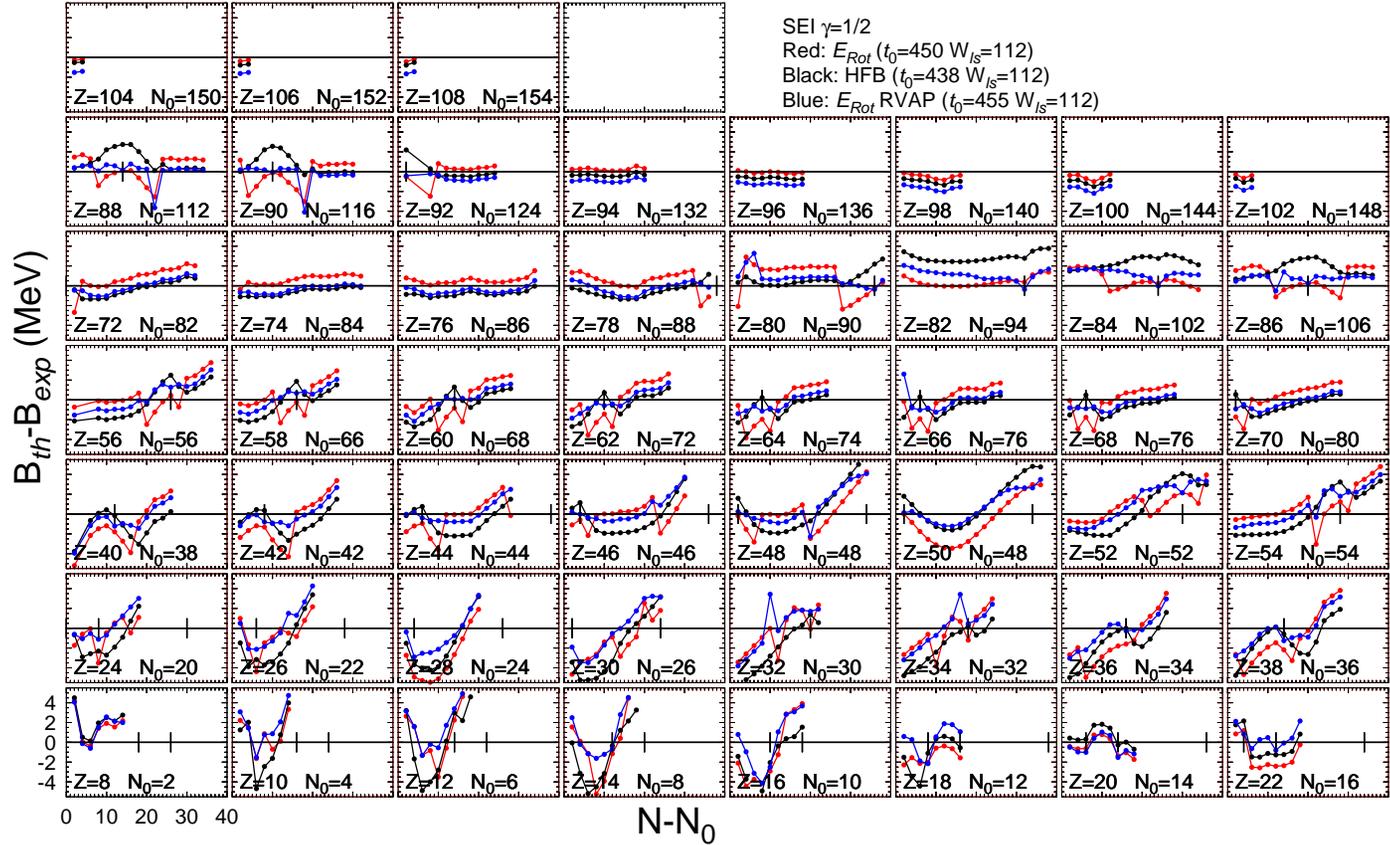}
\caption{(Color Online)The binding energy difference $\Delta B = B_\mathrm{th} - B_\mathrm{exp}$ 
computed with $\gamma=1/2$ is 
plotted as a function of the shifted (by $N_0$) 
neutron number $N - N_0$ for different isotopic chains. The 
proton number $Z$ and the neutron number 
shift $N_0$ are given in each panel. The ordinates range from -5.5 to 5.5 MeV with long ticks every 1 MeV. 
The $N - N_0$ axis extends over a range of 40 units with long ticks every 10 units and short ticks every 1 unit.
The horizontal line at $\Delta B=0$ is only for guiding the eye. Perpendicular marks indicate the position 
of the neutron magic numbers.}
\label{DeltaE}
\end{sidewaysfigure}

\begin{sidewaysfigure}
\vspace*{15cm}
\centering
\includegraphics[width=0.45\columnwidth,angle=-90]{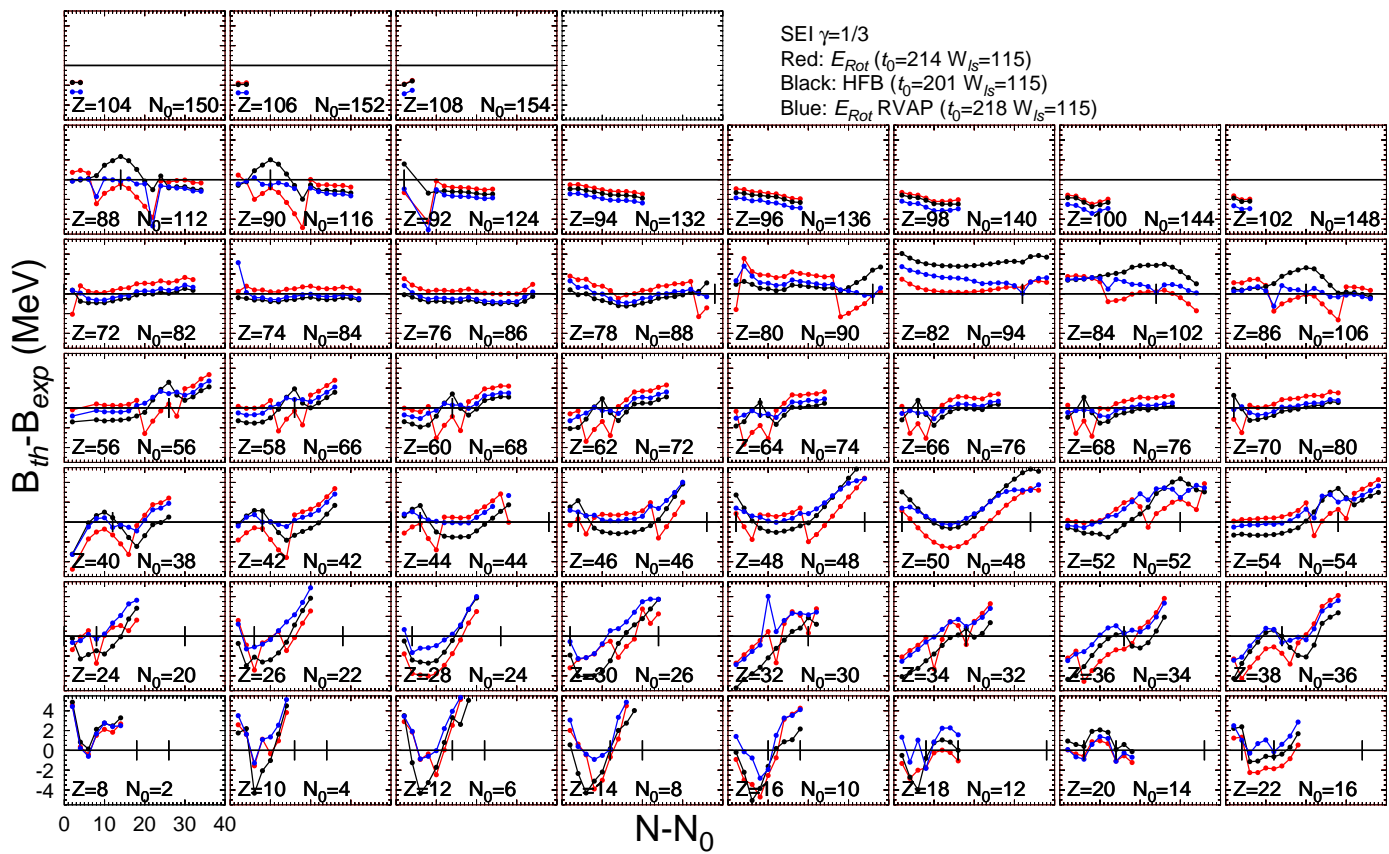}
\caption{(Color Online)The same as in figure \ref{DeltaE} but for the SEI corresponding to $\gamma=1/3$.}
\label{DeltaEp}
\end{sidewaysfigure}

The minimization process reveals that in all the cases the optimal 
values of $t_0$  and $W_0$ are basically the same as the values fitted 
in \cite{trr15}. 
These parameters,
as well as the parameter $x_0$ (see equation~(\ref{eq9a}))
together with
the minimal value of $\sigma(E)$ for the SEI sets having
$\gamma=1/3$ and $\gamma=1/2$,
are reported in Table \ref{tab:params}
for the three different methods of calculating the binding energy.
In Figures \ref{DeltaE} and \ref{DeltaEp} we plot the binding energy difference 
$\Delta B = B_\mathrm{th}-B_\mathrm{exp}$ for the $\gamma=1/2$ and $\gamma=1/3$ cases, respectively,
 as obtained with the three methods mentioned above.
 The results are shown for several isotopic chains ranging from $Z=8$ to $Z=108$.
The three curves displayed in each isotopic chain correspond to the HFB energy (black curve), the HFB energy corrected with the rotational energy in the projection after variation (PAV) way (red curve) and the HFB energy corrected with the
rotational energy but in the spirit of the
Restricted Variation after Projection (RVAP) approach (blue curve). The quantity
$\Delta B$ is plotted as a function of the neutron number $N$ shifted
by $N_0$ units, where $N_0$ indicates the origin of the horizontal axis for the
different isotopic chains displayed in these figures. 
Perpendicular marks in the $\Delta B=0$ horizontal lines indicate the position of 
neutron magic nuclei. Globally, these figures look qualitatively
similar to the figure displayed in \cite{BCPM13} computed using the BCPM 
energy density functional where the same color indexing was used 
(black for pure HFB, red for HFB with PAV and blue for HFB with RVAP). 
As mentioned, we have adjusted the parameters $t_0$ and 
$W_0$ to minimize $\sigma(E)$  in the three cases. 
It is found as a general feature that there is a degradation in 
the agreement with 
experimental data for light nuclei with $Z$ values less than 50.
For $Z> 50$ the behavior of $\Delta B$ is almost 
flat, whereas below $Z=50$ it behaves approximately as a straight 
line with a large slope. For light nuclei the curves show a more 
erratic behavior and the agreement between the theoretical predictions 
and the experimental values worsens (the differences between theory and experiment may be as large as $\pm$ 4 MeV in some particular cases). This lack of
accuracy in describing the experimental masses in light nuclei points
out the fact that, in general, light nuclei are not very well described 
at the mean field level ( also see in this respect Figure 5 of
Ref.~\cite{BCPM13}). For the HFB result we can also mention the 
significant failure to reproduce isotopic chains with magic $Z$ values. 
A typical example is Pb 
(and Po and Rn) where $\Delta B$ values as 
large as 3.5 MeV are obtained for most of the nuclei in the chain. This 
is in strong contrast with other isotopic chains  like the ones of Os 
and W where an almost perfect agreement with experiment is observed.
Peaks in $\Delta B$ are observed also for magic neutron numbers.
The rotational energy correction (PAV) improves the situation around Pb,
but the price to pay are the jumps observed in many isotopic chains
around magic neutron numbers. These jumps are due to the sudden transition from
spherical to deformed configurations, with the associated increase in
$E_\mathrm{rot}$ from zero to a couple
of MeV. The inclusion of the rotational correction
improves a little the rms deviation $\sigma (E)$ between the theoretical and experimental binding energies, but not in a significant way.
Finally, the RVAP correction smooths out the $\Delta B$ curves and 
reduces the $\sigma (E)$ value by almost 300 keV in the $\gamma=1/2$ 
case. One might wonder at this point about other corrections to the HFB 
energy that could help to improve the $\sigma (E)$ value. Obvious 
candidates are the correlation energy associated to particle number 
restoration, or the zero point energy corrections of quadrupole and 
octupole motion. The octupole zero point energy was investigated in
\cite{Rob15} for the Gogny force and the conclusion was that it played 
a minor role in improving $\Delta B$. Work to evaluate the other two 
corrections is in progress.

\begin{table*}
\caption{Values of the parameters $t_0$, $x_0$ and $W_0$, where $x_0$ is determined 
from equation~(\ref{eq9a}),
for the two EOSs corresponding to $\gamma=$1/3 and 1/2.
The {\it rms} deviations with respect to experiment in the binding 
energies of 620 even-even nuclei and charge radii of 313 even-even nuclei 
are also shown. The column $E_\mathrm{rot}$ indicates what kind of
rotational correction is included in the calculation.
}
\begin{center}
\begin{tabular}{ccccccc}\hline \hline
$\gamma$&$t_0$&$x_0$&$W_0$&$\sigma (E) $&$\sigma(R)$&$E_\mathrm{rot}$ \\
&MeV fm$^3$&&MeV&MeV&fm \\ \hline
1/3 & 201 & 3.1931 & 115 & 1.873 &  0.0253   & no \\
1/2 & 438 & 1.4182 & 112 & 1.958 &  0.0252   & no \\
1/3 & 214 & 3.0595 & 115 & 1.788 & 0.0253    &  yes: VAP \\
1/2 & 450 & 1.4071 & 112 & 1.843 & 0.0252    &  yes: VAP \\
1/3 & 218 & 3.0221 & 115 & 1.742 & 0.0255    &  yes: RVAP \\
1/2 & 455 & 1.4027 & 112 & 1.561 & 0.0255    &  yes: RVAP \\
\hline \hline
\end{tabular}
\end{center}
\label{tab:params}
\end{table*}

We can compare the rms deviations $\sigma(E)$ obtained using the SEI with the predictions of Gogny forces
for the same set of nuclei evaluated in the same conditions. At pure HFB level the $\sigma(E)$ values for the Gogny
D1S, D1N and D1M are 3.48, 4.88 and 5.08 MeV, respectively. If the rotational 
energy correction is added, the binding energy {\it rms} deviations reduce to 2.15 (D1S), 2.84 (D1N) and 
2.96 (D1M) MeV. However, if an additional global shift in the binding energy is added, one obtains $\sigma(E)$ values
of 2.14 (D1S), 1.47 (D1N) and 1.45 (D1M). As explained in \cite{BCPM13}, this global shift has been 
included to simulate the zero point quadrupole energy correction included
in the fitting protocol of D1M. 
Therefore, these reductions in the {\it rms} deviations in D1S, D1N and D1M are not surprising.

We have also explored the role of pairing correlations in the $\sigma (E)$ values
by multiplying the pairing strengths for protons and neutrons by 
factors $f_p$ and $f_n$, respectively, taking the values 0.95 and 1.05 (a 5\%
variation in the pairing strength). We have kept the same $t_0$ and
$W_0$ values as in the  $f_p=f_n=1$ case, as an
unrestricted search including four parameters would be computationally
expensive. The results show that the $f_p=f_n=1$ values
provide the optimal value of $\sigma (E)$.

\begin{figure}
\vspace{0.6cm}
\begin{center}
\includegraphics[width=0.5\textwidth,angle=-90]{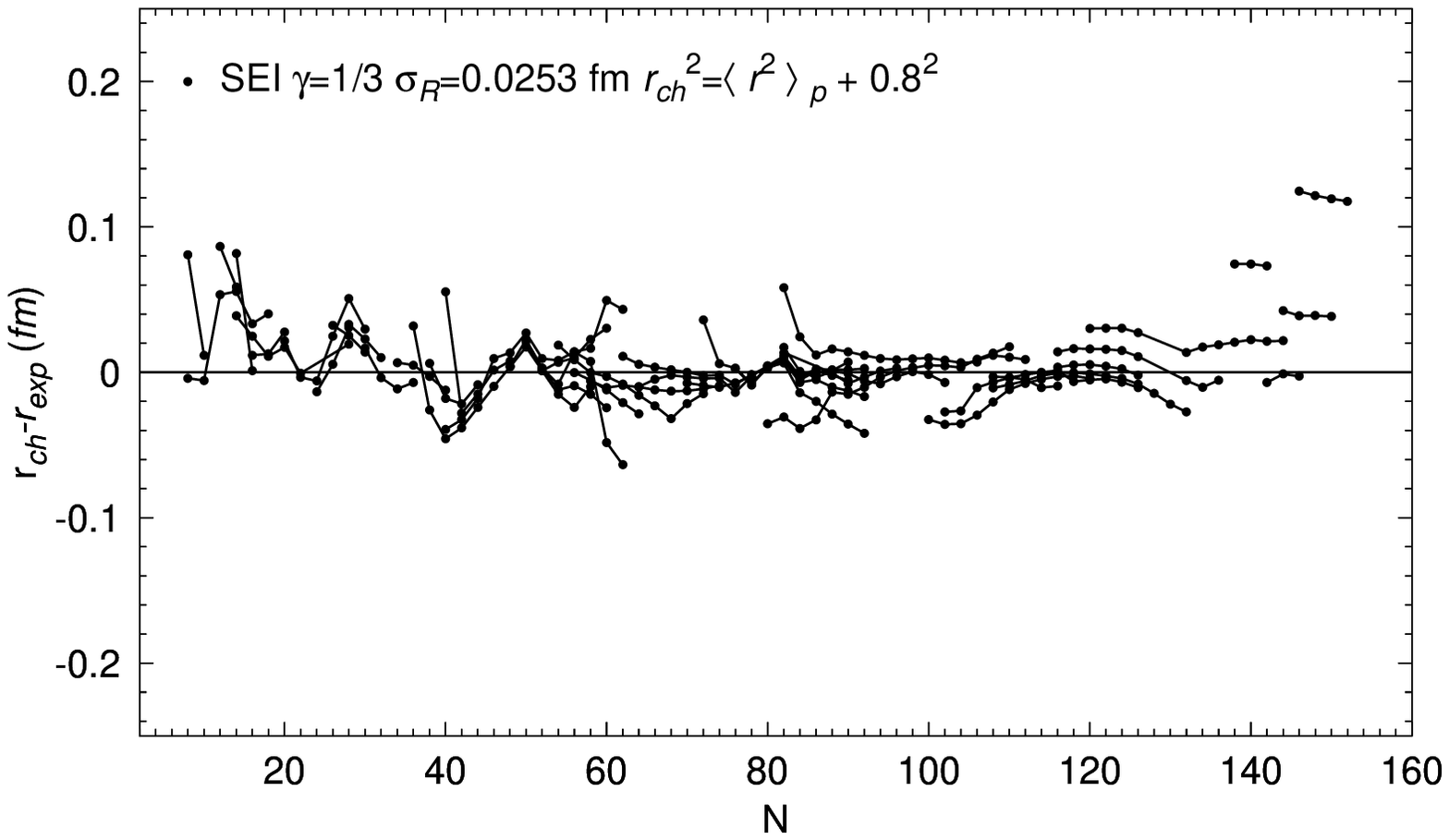}
\caption{The $\Delta r=r_\mathrm{ch}-r_\mathrm{exp}$ deviation is plotted as a
function of the neutron number $N$ 
for the $\gamma=1/3$ EOS of SEI.
}
\label{DeltaR}
\end{center}
\end{figure}

The nuclear charge radius is a relevant observable connected with the size of nuclei. It can be measured, for instance,
through electron scattering experiments. In our calculation the charge radii are estimated from the point-like 
proton distributions taking into account a proton radius of 0.8 fm \cite{amsler08}. The theoretical predictions
of SEI with $\gamma=1/3$ and $\gamma=1/2$ are compared to the measured values of the charge radii of 313 even-even nuclei   
reported in the Angeli's compilation \cite{angeli04}. In our previous study of spherical nuclei \cite{trr15},
where we considered a reduced set of 86 even-even nuclei, we obtained {\it rms} deviations $\sigma(R)$ in the charge radius of
0.017 fm and 0.016 fm for $\gamma=1/3$ and $\gamma=1/2$, respectively. In the present calculation, 
considering 313 spherical and deformed nuclei, we obtain $\sigma(R)$=  0.0253 fm and 0.0252 fm for the two parameter sets, respectively.
These values can be compared with the rms deviations in charge radii obtained for Gogny forces for the same set of nuclei, which are 
0.037 fm for D1S and 0.028 fm
for D1M.  The differences $\Delta r=r_\mathrm{ch}-r_\mathrm{exp}$ are shown
in Figure \ref{DeltaR} as a function of mass number and for the different isotopic
chains. We observe a reasonable agreement with experiment with all the points
scattered around the zero line except for light nuclei and some very heavy
ones. The departure of our predictions for the very heavy nuclei is not in agreement
with the overall behavior and deserves further consideration.

\subsection{Comparison with previous calculations of spherical nuclei using the SEI}

The non-local Density Functional Theory has been discussed in earlier 
literature (see e.g. \cite{vinas03} and references therein). As it was 
shown in \cite{vinas03}, the Lieb theorem  \cite{lieb83}, which 
establishes the many-to-one mapping of the $A$ particle Slater 
determinant wave-functions onto the local particle density, allows to 
write the energy density functional in the non-local case as 
$\varepsilon[\tilde{\rho_0}] = \varepsilon_0[\tilde{\rho_0}] + 
E_{RC}[\rho]$. In this expression $\tilde{\rho_0}$ is a Slater 
determinant density matrix associated to an effective Hamiltonian 
$\tilde{H}$, $\varepsilon_0[\tilde{\rho_0}]$ is the corresponding 
Hartree-Fock (HF) energy and the remaining part $E_{RC}[\rho]$, which 
is a function of the local density $\rho$ only, is the residual 
correlation energy. Effective forces of Gogny, M3Y and SEI type contain 
a finite range part plus a density dependent contribution. The
HF energy density associated to these forces can be regarded as a
possible realization of the exact energy density functional because the 
exact correlation energy density is in general unknown and it is 
approximated by the density dependent term of these finite-range 
effective forces. In Ref.~\cite{vinas03} it is also shown that another 
reduction can be done by mapping the Slater determinant density matrix 
onto a set $\rho^{QL}$ of local particle, kinetic energy and spin 
densities, which allows to write the energy density in a quasi-local
form $\varepsilon[\rho^{QL}] = \varepsilon_0[\rho^{QL}]
+ E_{RC}[\rho]$. In Ref.~\cite{trr13}, as discussed in \cite{vinas03},
we performed an additional approximation by computing the quasi-local 
energy density corresponding to the exchange term using the Extended 
Thomas-Fermi expansion of the density matrix \cite{vinas00}. This 
approximation is similar to the density matrix expansions
 proposed by Negele and Vautherin \cite{negele72} and Campi and Bouyssy \cite{campi78}
(see Ref.~\cite{vinas00} for further details). The variational principle applied to this 
quasi-local energy density functional allows to write formally the equations of motion in a 
similar way to those obtained with the Skyrme forces \cite{vinas03}. As a consequence,
at least for spherical nuclei, the calculations can be performed in coordinate space
avoiding the expansion of the wave-functions in a basis.  
  
\begin{table}
\caption{Comparison of the quantal binding energies and radii obtained in doubly magic nuclei
using the QLEDF of \cite{trr13} and the HF approach.
\\
}
\renewcommand{\tabcolsep}{0.45cm}
\renewcommand{\arraystretch}{1.2}
\begin{tabular}{cccccccccccccccc}
\hline
\hline
Nucleus & $E_{EDF}$(MeV) & $E_{HF}$(MeV) & $r_{p}^{EDF}$(fm) & $r_{p}^{HF}$(fm) \\
\hline
\hline
$^{16}$O   & -127.6240 & -127.0907 & 2.6594 & 2.6646\\
$^{28}$O   & -179.4020 & -178.5816 & 2.7832 & 2.7922\\
$^{40}$Ca  & -342.1981 & -341.2690 & 3.3997 & 3.4053\\
$^{48}$Ca  & -416.8068 & -414.6866 & 3.4210 & 3.4326\\
$^{56}$Ni  & -478.7936 & -480.0659 & 3.6847 & 3.6907\\
$^{78}$Ni  & -645.4516 & -646.4948 & 3.8814 & 3.8928\\ 
$^{100}$Sn & -824.8094 & -825.6940 & 4.4132 & 4.4244\\
$^{132}$Sn &-1105.0788 &-1105.4751 & 4.6360 & 4.6456\\
$^{208}$Pb &-1636.6551 &-1635.8961 & 5.4285 & 5.4344\\ 
\hline
\end{tabular}
\label{tab:magic}
\end{table}

\begin{table}
\caption{Comparison of quantal binding energies and radii obtained in the 
isotopes of $Sn$
using the QLEDF plus improved BCS pairing of Ref.~\cite{trr13} and the
HFB method.
}
\renewcommand{\tabcolsep}{0.45cm}
\renewcommand{\arraystretch}{1.2}
\begin{tabular}{cccccccccccccccc}
\hline
\hline
Nucleus & $E_{EDF}$(MeV) & $E_{HFB}$(MeV) & $r_{p}^{EDF}$(fm) & $r_{p}^{HFB}$(fm) \\
\hline
\hline
$^{102}$Sn  & -848.1597 & -848.8267 & 4.4297 & 4.4389\\
$^{104}$Sn  & -870.2672 & -870.8571 & 4.4450 & 4.4535\\
$^{106}$Sn  & -891.6288 & -891.9747 & 4.4603 & 4.4685\\
$^{108}$Sn  & -912.2156 & -912.2815 & 4.4758 & 4.4839\\
$^{110}$Sn  & -932.3108 & -931.8436 & 4.4915 & 4.4995\\
$^{112}$Sn  & -951.4225 & -950.7060 & 4.5069 & 4.5150\\
$^{114}$Sn  & -969.8919 & -968.9010 & 4.5215 & 4.5301\\
$^{116}$Sn  & -987.5958 & -986.4534 & 4.5355 & 4.5447\\
$^{118}$Sn  &-1004.5842 &-1003.3858 & 4.5488 & 4.5586\\
$^{120}$Sn  &-1020.8849 &-1019.7195 & 4.5620 & 4.5721\\
$^{122}$Sn  &-1036.5196 &-1035.4725 & 4.5748 & 4.5850\\
$^{124}$Sn  &-1051.5581 &-1050.6561 & 4.5875 & 4.5977\\
$^{126}$Sn  &-1065.9662 &-1065.2716 & 4.6021 & 4.6100\\
$^{128}$Sn  &-1079.7420 &-1079.3066 & 4.6137 & 4.6221\\
$^{130}$Sn  &-1092.7420 &-1091.7298 & 4.6251 & 4.6341\\
\hline
\end{tabular}
\label{tab:Sn}
\end{table}

Table \ref{tab:magic} compares the binding energies and proton radii of some
doubly closed shell nuclei computed with the QLEDF
(see \cite{trr13} for further details) with the results of the full HF
method using the SEI with $\gamma=1/3$. From this table we see that the
quasi-local energy density functional approach predicts values that are
very close to the HF ones. The largest differences in binding energies
and proton radii are always less than 0.5\% and 0.3\%, respectively.
Pairing correlations have been included in Density Functional Theory
since long ago \cite{oliveira88,lathi04,luders05} and in the context of
the quasi-local reduction of the non-local theory in \cite{krewald06}.
Although HFB is the standard theory for dealing with pairing
correlations in finite nuclei, the simpler BCS method is often
sufficient to describe ground-state energies near the $\beta$-stability
valley \cite{bender00}. In Refs.~\cite{trr13} and \cite{trr15} we have
used  the QLEDF together with an improved BCS
approach \cite{patra01} to study the properties of finite nuclei. In Table
\ref{tab:Sn} we report the binding energies and proton radii for the Sn
isotopic chain from $N=50$ to $N=82$ computed in this way using the SEI
with $\gamma=1/3$ as well as the same quantities obtained using the
HFB method with the same interaction. From this Table we see again
that the quasi-local results lie close to the HFB ones and that the
largest differences, about 0.1\%, correspond to the mid-shell nuclei.
We have checked that similar results are obtained using the SEI with
$\gamma=1/2$ instead of  $\gamma=1/3$.
From the discussion in this subsection we can conclude that the quantal predictions
for spherical nuclei reported in Refs.~\cite{trr13} and \cite{trr15} by using the
QLEDF including pairing correlations, treated with the improved BCS approach
of Ref.~\cite{patra01}, are in excellent agreement
with the HFB results. A similar conclusion was reached in Refs.~\cite{vinas03} 
and \cite{krewald06} for Gogny forces.

\section{Deformation properties}

\subsection{Potential Energy Surfaces}

\begin{figure}
\vspace{0.6cm}
\begin{center}
\includegraphics[width=0.8\columnwidth,clip=true]{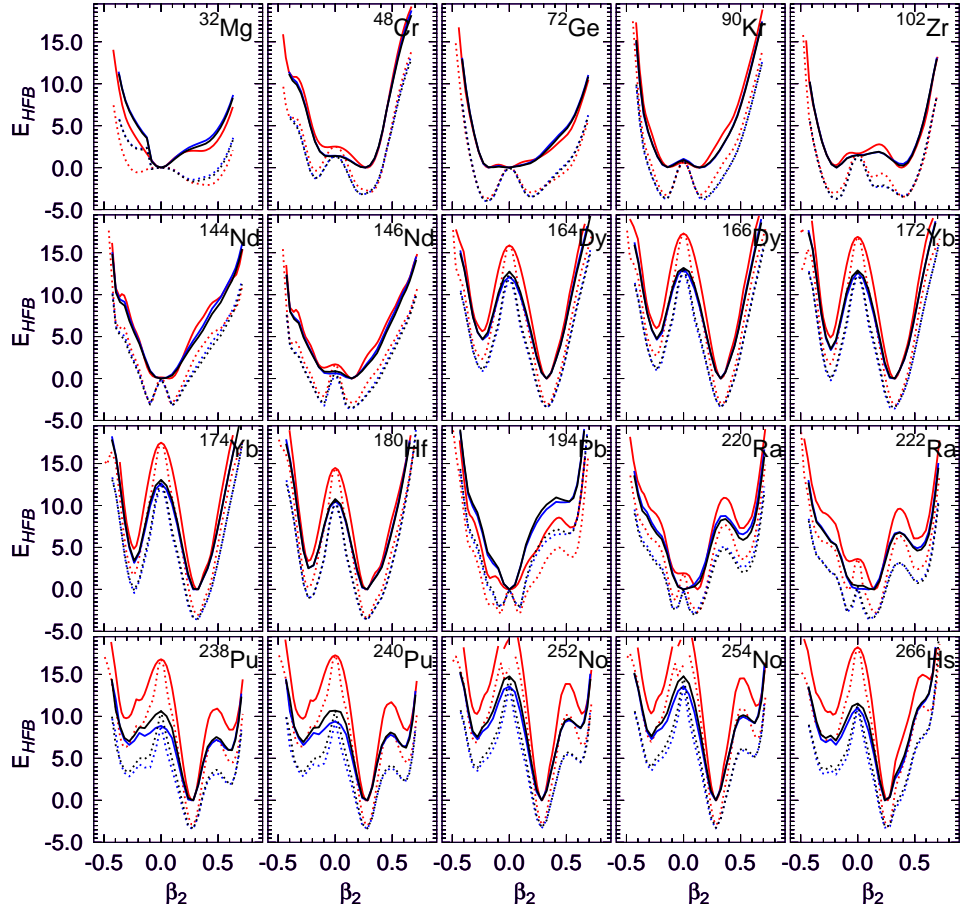}
\caption{(Color Online)Potential Energy Surface for selected deformed nuclei along the 
whole periodic table. Full (dotted) lines are used to represent the HFB
(HFB plus rotational correction) curves. Black and blue curves are for
SEI with $\gamma=1/2$ and $\gamma=1/3$, respectively, whereas the red curve
is for the Gogny D1S force. The HFB energies have been shifted as to make
the curves of the three forces coincide at the ground-state energy.}
\label{Fig. 5}
\end{center}
\end{figure}

The potential energy surface (PES), defined as the HFB energy computed 
with HFB wave functions constrained to definite values of the axial quadrupole 
moment, are plotted in Figure \ref{Fig. 5} 
for selected nuclei from light to very 
heavy ones. The Gogny D1S curves are given as a reference. The main 
conclusion is that the position of the maxima and minima obtained with the
two versions of SEI and Gogny D1S 
are located essentially at the same values of the 
quadrupole deformation. However, the spherical barrier (the energy difference
between the spherical maximum and the prolate ground state) is much 
lower in energy for the SEI 
interactions than for the Gogny force. Also the excitation energy of fission isomers or 
super-deformed states is slightly different depending on the forces 
used. This energy difference is strongly influenced by the position of 
the spherical single particle levels, which depend upon the average mean
field and the spin-orbit interaction. Finally, the barriers separating
the ground state minimum from the fission isomer are lower in SEI than
in Gogny D1S. In order to quantify the impact on observables due to the 
PES differences, a dynamical calculation of the Generator Coordinate 
Method (GCM) type would be required. This is the subject of active
research. 

\subsection{Fission Barriers}

\begin{figure}
\vspace{0.6cm}
\begin{center}
\includegraphics[width=0.8\columnwidth,clip=true]{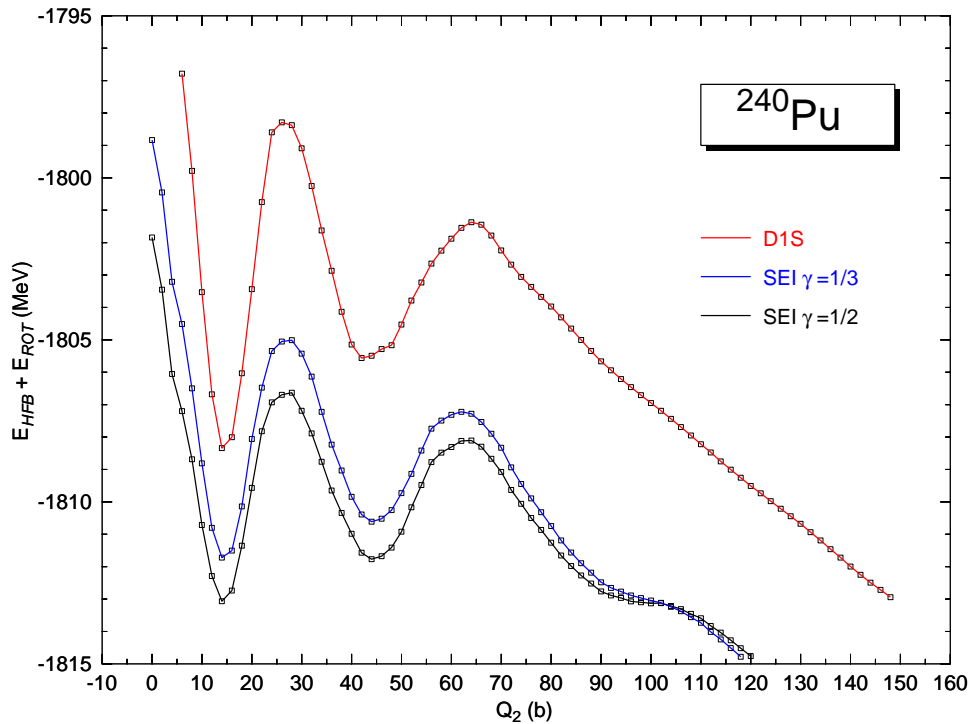}
\caption{(Color Online)Fission properties of $^{240}$Pu obtained with the SEI with
$\gamma=1/3$ and $\gamma=1/2$ are compared with the predictions of the D1S Gogny force.
The results for the HFB energy plus the rotational correction E$_\mathrm{rot}$ are depicted
as a function of the axial quadrupole moment for the three forces considered.
}
\label{Fig. 3}
\end{center}
\end{figure}

\begin{figure}
\vspace{0.6cm}
\begin{center}
\includegraphics[width=0.8\columnwidth,clip=true]{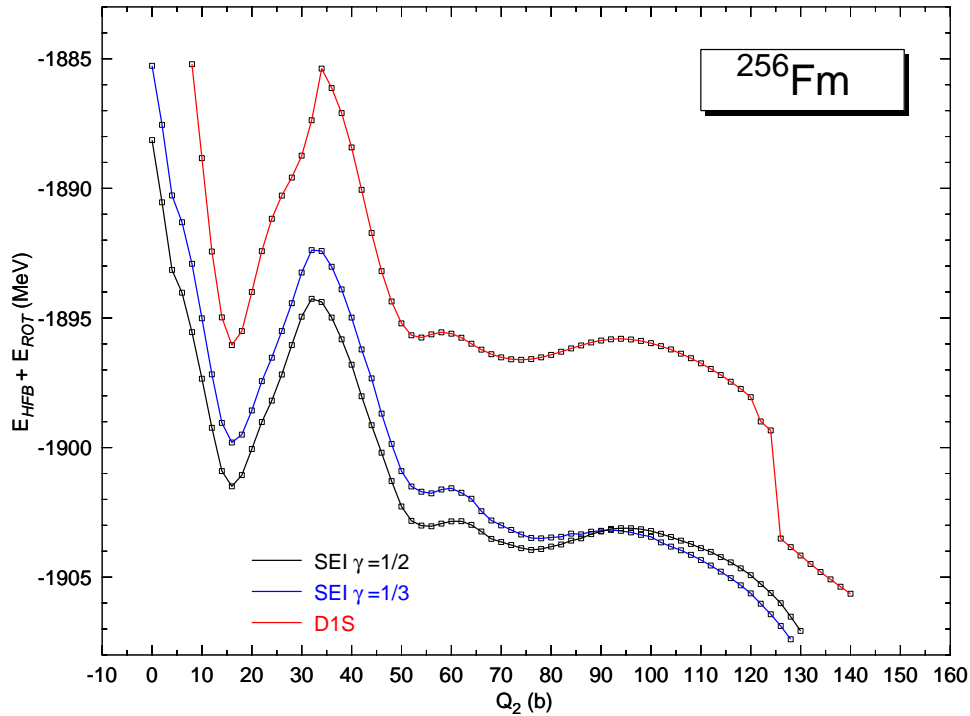}
\caption{(Color Online)The same as in Figure \ref{Fig. 3} for the nucleus $^{256}$Fm.}
\label{Fig. 4}
\end{center}
\end{figure}

Fission is one of the most characteristic aspects of nuclear dynamics. The description
of the dynamical evolution of the parent nucleus into two fragments is a real
challenge both for the quantum many body techniques to be used as well as for
the effective interactions. Therefore, it is a very good testing ground for
any newly proposed interaction. To test the performance of SEI in this matter,
we have followed the traditional approach \cite{Baran.15} and carried out 
constrained  mean field calculations with the mass quadrupole moment used as
the driving coordinate. The resulting PES obtained for the nuclei
$^{240}$Pu and $^{256}$Fm are shown, as an example, in Fig. \ref{Fig. 3} where
the HFB energy supplemented by the rotational correction, is plotted as 
a function of the quadrupole moment for the two sets of EOSs of SEI considered in this 
paper, as well as for  the Gogny D1S force. The Gogny D1S results are 
used here as a benchmark, given the fact that the force was adjusted
to fission barrier heights, and has been successfully tested with very 
different kinds of fission---see Ref \cite{Warda.11} as
an example. By looking at the PES in Fig. \ref{Fig. 3} we notice that
the quadrupole moment  of all the maxima and minima present in the PES is
roughly the same in the
three cases,  but the  energies, relative to the ground state minimum, change depending on the
interaction considered. For $^{240}$Pu, the SEI values for the first  
fission barrier height (energy difference between the first maximum
at around $Q_2=25$ fm$^2$ and the ground state minimum) are smaller 
than the one obtained with Gogny D1S. However, the SEI values for the first 
fission barrier height are very close in both cases to the accepted experimental value 
of 6.05 MeV \cite{Ber15}. The trend is similar for the second barrier,
located at around $Q_2=65$ fm$^2$, with heights (measured with respect to
the ground state energy) in the SEI case lower than the height obtained for Gogny D1S.
The SEI values are a few hundred keV lower than the experimental value of 5.15 MeV \cite{Ber15} of the second barrier. 
Finally, the excitation energy of the fission isomer (local minimum at 
around $Q_2=45$ fm$^2$) obtained with both SEI forces is lower
than the Gogny prediction and around 1.5 MeV  lower than the experimental value of 
2.8 MeV. In spite of these small discrepancies we can consider that the 
fission properties of SEI for $^{240}$Pu are in reasonable agreement with 
the ones of the Gogny force as well as with the experimental data. In the case of 
$^{256}$Fm, there is no experimental data to compare with. The agreement with 
the Gogny D1S curve is remarkable until the end of the first barrier 
but from there on, it deviates a little bit as a consequence of different shell 
effects. In both cases, a dynamical treatment including the quadrupole 
collective inertia and  zero point energy corrections would be required
for the calculation of the spontaneous fission life-time. Work along this
direction is in progress and will be reported in the future.

\section*{Conclusions}

We have performed a study of finite nuclei using the SEI including 
deformation degrees of freedom. This study extends the results for 
spherical nuclei reported in previous literature \cite{trr13,trr15}. In 
this work we have computed ground state properties of deformed nuclei 
using the Hartree-Fock-Bogoliubov method with a harmonic oscillator 
basis. This calculation of deformed nuclei includes the rotational 
energy correction, computed using the restricted variation after 
projection method. The SEI, with the parameters fitted as explained in 
this work, is able to describe the binding energies of 620 and the 
charge radii of 313 even-even nuclei with {\it rms} deviations about 
1.8 MeV and 0.025 fm, respectively. These deviations are similar to 
those found, for the same set of nuclei and computed in the same 
conditions, using the D1S, D1N and D1M Gogny forces.

Concerning deformation properties predicted by the SEI, it is found 
that the Potential Energy Surfaces along the whole periodic table 
follow quite well the pattern obtained using the D1S Gogny force. 
Nevertheless, some differences appear in the comparison between the SEI and
the D1S results. In particular, the spherical maxima and super-deformed minima 
predicted in the potential energy surfaces with the SEI models are systematically 
lower than the corresponding ones obtained with the Gogny D1S force. 
This suggests that the surface energy parameter of both SEI models is smaller
than the one of the D1S force. 
We have also studied the fission barriers of
the nuclei $^{240}$Pu and $^{256}$Fm with the SEI. For the nucleus
$^{240}$Pu this  model predicts around (slightly depending on $\gamma$)
6.0 MeV for the first  fission barrier, around 4.5 MeV for the second fission barrier and an excitation energy of the fission isomer of around 1.5 MeV. These values
agree quite well with the accepted experimental values \cite{Ber15}.

The present work confirms, as suggested by our previous analysis in 
\cite{trr13} and \cite{trr15}, the ability of the SEI to reproduce 
properties of different types of nuclear matter and, simultaneously, 
predict finite nuclei properties with similar quality
to well-known effective interactions. At variance with other 
commonly used effective forces, such as Skyrme, Gogny or M3Y, ten of 
the twelve parameters of the SEI  are determined using 
experimental/empirical constraints in nuclear matter as well as
to reproduce the behavior predicted by microscopic calculations 
in symmetric nuclear 
matter and spin polarized neutron matter. In particular, the momentum 
dependence of the mean field derived from the interaction is fitted in 
such a way that the effective mass splitting in both, asymmetric 
nuclear matter and spin polarized neutron matter, reproduce the results 
provided by microscopic Dirac-Brueckner-Hartree-Fock calculations. 
We note that only the parameter $t_0$ of SEI and the strength of the
spin-obit interaction $W_0$ are fitted to finite nuclei. Therefore, the
results for finite nuclei properties obtained with the SEI are,
actually, predictions of the model. We have focused our study on 
even-even nuclei. To extend the study to odd-even, even-odd and 
odd-odd nuclei requires to develop completely polarized asymmetric 
nuclear matter in order to deal with odd components in the energy 
density. Work in this direction is left for a future contribution. 

\section*{Acknowledgments}

One author (TRR) thanks the Departament d'Estructura i Constituents de 
la Mat\`eria, University of Barcelona, Spain for hospitality during the 
visit. The work is covered under FIST programme of School of Physics, 
Sambalpur University, India. Authors M.C. and X.V. acknowledge partial support from 
Grant No. FIS2014-54672-P from Spanish MINECO and FEDER, Grant No. 2014SGR-401 
from Generalitat de Catalunya, the Consolider-Ingenio Programme CPAN 
CSD2007-00042, and the project MDM-2014-0369 of ICCUB (Unidad de Excelencia 
Mar\'{\i}a de Maeztu) from MINECO.

\section*{Appendix A}

As mentioned in section 2, asymmetric nuclear matter computed with the SEI 
in equation (\ref{eq1}) depends on the parameters $\gamma$, $b$, 
$\varepsilon_{0}^{l}$, $\varepsilon_{0}^{ul}$, 
$\varepsilon_{\gamma}^{l}$,$\varepsilon_{\gamma}^{ul}$, $\varepsilon_{ex}^{l}$,
 $\varepsilon_{ex}^{ul}$ and $\alpha$. 
The new parameters are connected to the parameters of the interaction
through the relations,
\begin{eqnarray}
\varepsilon_{0}^{l}=\rho_0\left[\frac{t_0}{2}\left(1-x_0\right)
+\left(W+\frac{B}{2}-H-\frac{M}{2}\right)\pi^{3/2}\alpha^3\right] \nonumber \\
\nonumber \\
\varepsilon_{0}^{ul}=\rho_0\left[\frac{t_0}{2}\left(2+x_0\right)
+\left(W+\frac{B}{2}\right)\pi^{3/2}\alpha^3\right] \nonumber \\
\nonumber \\
\varepsilon_{\gamma}^{l}=\frac{t_3}{12}\rho_0^{\gamma+1}(1-x_3),
\varepsilon_{\gamma}^{ul}=\frac{t_3}{12}\rho_0^{\gamma+1}(2+x_3) \nonumber \\
\nonumber \\
\varepsilon_{ex}^{l}=\rho_0\left(M+\frac{H}{2}-B-\frac{W}{2}\right)
\pi^{3/2}\alpha^3 \nonumber \\
\nonumber \\
\varepsilon_{ex}^{ul}=\rho_0\left(M+\frac{H}{2}\right)
\pi^{3/2}\alpha^3.
\end{eqnarray}

The energy density in asymmetric nuclear matter reads 
\begin{eqnarray}
H(\rho_n,\rho_p)&=&\frac{3\hslash^2}{10m}\left(k_n^2\rho_n+k_p^2\rho_p\right)
+\frac{\varepsilon_{0}^{l}}{2\rho_0}\left(\rho_n^2+\rho_p^2\right)
+\frac{\varepsilon_{0}^{ul}}{\rho_0}\rho_n\rho_p \nonumber \\
&&+\left[\frac{\varepsilon_{\gamma}^{l}}{2\rho_0^{\gamma+1}}\left(\rho_n^2+\rho_p^2\right)
+\frac{\varepsilon_{\gamma}^{ul}}{\rho_0^{\gamma+1}}\rho_n\rho_p\right]
\left(\frac{\rho({\bf R})}{1+b\rho({\bf R})}\right)^{\gamma} \nonumber \\
&&+\frac{\varepsilon_{ex}^{l}}{2\rho_0}(\rho_n^2 J(k_n) + \rho_p^2 J(k_p))\nonumber \\
&&+\frac{\varepsilon_{ex}^{ul}}{2\rho_0}\frac{1}{\pi^2}\left[\rho_n\int_0^{k_p}I(k,k_n)k^2dk
+\rho_p\int_0^{k_n}I(k,k_p)k^2dk\right] \nonumber \\
\end{eqnarray}
where, the functions $J(k_i)$ and $I(k,k_i)$ with $k_i=(3 \pi^2 \rho_i)^{1/3}$
($i=n,p$) are given by
\begin{eqnarray}
J(k_i)= \frac{3\Lambda^3}{2{k_i^3}} \bigg[\frac{\Lambda^3}{8{k_i^3}}
-\frac{3 \Lambda}{4{k_i}}
 -\left(\frac{\Lambda^3}{8{k_i^3}}-\frac{\Lambda}{4{k_i}}\right) e^{-4{k_i^2}/\Lambda^2}
+\frac{\sqrt{\pi}}{2} \textrm{erf}\left(2k_i/\Lambda\right) \bigg]
\label{eq10a}
\end{eqnarray}
and
\begin{eqnarray}
I(k,k_{i}) &=& \frac{3 \Lambda^3}{8k_{i}^{3}}
\bigg[\frac{\Lambda}{k}
\left(e^{-\left(\frac{k+k_{i}}{\Lambda}\right)^2}
-e^{-\left(\frac{k-k_{i}}{\Lambda}\right)^2}\right) \nonumber \\
&+&
\sqrt{\pi}\left(\textrm{erf}{\bigg(\frac{k+k_{i}}
{\Lambda}\bigg)}-\textrm{erf}{\bigg(\frac{k-k_{i}}
{\Lambda}\bigg)}\right) \bigg] ,
\label{ppnma}
\end{eqnarray}
with $\Lambda=2/\alpha$.

The energy density in polarized pure neutron matter is given in \cite{trr15}.
In principle, the parameters $\varepsilon_{0}^{l}$, $\varepsilon_{\gamma}^{l}$ and
$ \varepsilon_{ex}^{l}$ split into the interactions between pairs of neutrons
with the same (index ${l,l}$) and the opposite (index ${l,ul}$) spin orientation:
\begin{eqnarray} \nonumber
H_{pol}^N(\rho_{nu},\rho_{nd}) &=&
\bigg[ \frac{3\hslash^{2}(k_{nu}^{2}\rho_{nu}+k_{nd}^{2}\rho_{nd})}{10M}
+\frac{\varepsilon_{0}^{l,l}}{2\rho_0}(\rho_{nu}^{2}
+\rho_{nd}^{2}) \nonumber \\
&+&\frac{\varepsilon_{0}^{l,ul}}{\rho_0}\rho_{nu}\rho_{nd}
+ \bigg(\frac{\varepsilon_{\gamma}^{l,l}}{2\rho_0^{\gamma+1}}
(\rho_{nu}^{2}+\rho_{nd}^{2})+\frac{\varepsilon_{\gamma}^{l,ul}}
{\rho_0^{\gamma+1}}\rho_{nu}\rho_{nd} \bigg)\left(\frac{\rho({\bf R})}
{1+b\rho({\bf R})}\right)^{\gamma} \nonumber \\
&+&\frac{\varepsilon_{ex}^{l,l}}{2\rho_0} \bigg( \rho_{nu}^{2} J(k_{nu})
+\rho_{nd}^{2} J(k_{nd}) \bigg) \nonumber \\
&+&\frac{\varepsilon_{ex}^{l,ul}}{4\rho_0\pi^{2}} \bigg(
\rho_{nu} \int_0^{k_{nd}}\!\! I(k,k_{nu})k^{2}dk
+ \rho_{nd} \int_0^{k_{nu}}\!\! I(k,k_{nd})k^{2}dk \bigg) \bigg] .
\nonumber \\
\label{ppnm}
\end{eqnarray}
However, as explained in \cite{trr15} and mentioned before, the limit of fully
polarized neutron matter requires $\varepsilon_{\gamma}^{l,l}=0$ and
$\varepsilon_{0}^{l,l}=-\varepsilon_{ex}^{l,l}$ and the fit to the DBHF effective
mass in polarized neutron matter determines $\varepsilon_{ex}^{l,l}=\varepsilon_{ex}^{l}/3$

\section*{Appendix B}
 
The one-body density matrix as well as the particle, kinetic energy and spin local densities 
are obtained from the single-particle orbitals $\phi_i$ that define the Slater
determinant $\Psi_0$ as
\begin{eqnarray}
\rho_q ({\bf r,\bf r'})= \sum_{i=1}^{A_q} \sum_{\sigma} \phi_i^*({\bf r}, \sigma, q)
\phi_i({\bf r'}, \sigma, q), 
\end{eqnarray}
\begin{eqnarray}
\rho_q ({\bf r})= \sum_{i=1}^{A_q} \sum_{\sigma} \vert \phi_i({\bf r}, \sigma, q)\vert^2,
\end{eqnarray}
\begin{eqnarray}
\tau_q ({\bf r})= \sum_{i=1}^{A_q} \sum_{\sigma} \vert{\bf \triangledown}\phi_i
({\bf r}, \sigma, q)\vert^2,
\label{eq19}
\end{eqnarray}
and
\begin{eqnarray}
J_q ({\bf r})=i \sum_{i=1}^{A_q} \sum_{\sigma,\sigma'} \phi_i^* ({\bf r}, \sigma, q)
\left[({\bf \sigma})_{\sigma,\sigma'} \times{\bf \triangledown}\right] \phi_i
({\bf r}, \sigma, q),
\end{eqnarray} 
respectively.
The total energy density ${\mathcal H}$ can be written as
\begin{eqnarray}
{\mathcal H}&=&\frac{\hslash^2}{2m}\left(\tau_n+\tau_p\right)+
{\mathcal H}_{d}^{Nucl}+{\mathcal H}_{exch}^{Nucl}+{\mathcal H}_{zr}^{Nucl}
 + {\mathcal H}^{SO} +{\mathcal H}^{Coul},
\end{eqnarray}
where, we have split the total energy density as the sum of the kinetic part, direct, exchange and zero-range nuclear
terms together with the spin-orbit and Coulomb contributions.

The Coulomb energy is taken in the usual way as a direct plus Slater exchange contribution 
computed with the point proton density:
\begin{eqnarray}
{\mathcal H}^{Coul}({\bf R})= \frac{e^2}{2}\int\frac{\rho_p({\bf
R}+\frac{\bf r}{2})\rho_p({\bf R}-\frac{\bf r}{2})}
{r}d^3r
-\frac{3e^2}{4}\left(\frac{3}{\pi}\right)^{1/3}\rho_p^{4/3}({\bf R}),
\end{eqnarray}
where, ${\bf r=r_1-r_2}$ and ${\bf R=\frac{r_1+r_2}{2}}$ are the relative and
center of mass coordinates, respectively, for the two interacting nucleons
located at ${\bf r_1}$ and ${\bf r_2}$.

In the SEI model the spin-orbit interaction is chosen in the form used in the
case of Skyrme and Gogny forces: $v_{i,j}^{SO}=iW_0\left({\bf \sigma_i}+{\bf \sigma_j}\right)
\left[{\bf k'\times\delta(r_i,r_j)k}\right]$. The corresponding contribution to
the energy density is  
\begin{eqnarray}
{\mathcal H^{SO}}({\bf R}) = -\frac{1}{2}W_0\left[\rho({\bf R}){\bf \nabla J}
+ \rho_n({\bf R}){\bf \nabla J_n} + \rho_p({\bf R}){\bf \nabla J_p} \right].
\label{eq23}
\end{eqnarray}
The direct and exchange contributions to the nuclear energy density read
\begin{eqnarray}
&&{\mathcal H}_{d}^{Nucl}({\bf R})=\frac{1}{2}\int d^3r e^{-r^2/\alpha^2}\bigg[\left(W+\frac{B}{2}-H
-\frac{M}{2}\right)\rho({\bf R}+\frac{\bf r}{2})\rho({\bf R}-\frac{\bf r}{2}) \nonumber \\
&&- \left(W+\frac{B}{2}\right)\left(\rho_n ({\bf R}+\frac{\bf r}{2})
\rho_p({\bf R}-\frac{\bf r}{2}) +\rho_p({\bf R}+\frac{\bf r}{2})\rho_n({\bf
R}-\frac{\bf r}{2})\right)\bigg],
\end{eqnarray}
and
\begin{eqnarray}
{\mathcal H}_{exch}^{Nucl}({\bf R})&=&\int d^3r e^{-r^2/\alpha^2}\bigg[
\frac{1}{2}\left(M+\frac{H}{2}-B-\frac{W}{2}\right)
\big[\rho_n^2({\bf R}+\frac{\bf r}{2},{\bf R}-\frac{\bf r}{2}) 
\nonumber \\
&& +\rho_p^2({\bf R}+\frac{\bf r}{2},{\bf R}-\frac{\bf r}{2})\big]
\nonumber \\
&& + \left(M+\frac{H}{2}\right) 
\rho_n({\bf R}+\frac{\bf r}{2},{\bf R}-\frac{\bf r}{2})
\rho_p({\bf R}+\frac{\bf r}{2},{\bf R}-\frac{\bf r}{2})\bigg].
\label{eq18}
\end{eqnarray}
Finally, the contribution from the zero-range part of the interaction 
can be written as
\begin{eqnarray}
{\mathcal H}_{zr}^{Nucl} &=& \frac{t_0}{4}\left[(1-x_0)\left[\rho_n^2({\bf R})+\rho_p^2({\bf R})\right]
+(4+2x_0)\rho_n({\bf R})\rho_p({\bf R})\right] \nonumber \\
&& +\frac{t_3}{24}\left[(1-x_3)\left[\rho_n^2({\bf R})+\rho_p^2({\bf R})\right]\right]
\left(\frac{\rho({\bf R})}{1+b\rho({\bf R})}\right)^{\gamma} \nonumber \\
&&+ \frac{t_3}{24}\left[ (4+2x_3)\rho_n({\bf R})\rho_p({\bf R})\right]
\left(\frac{\rho({\bf R})} {1+b\rho({\bf R})}\right)^{\gamma} .
\end{eqnarray}

\end{document}